# Effect of the Vacuum Energy Density on Graviton Propagation

Giovanni Modanese[1] and Giorgio Fontana[2]

[1]Logistics and Production Engineering, University of Bolzano, Via Sernesi 1, I-39100 Bolzano, Italy
[2]University of Trento, I-38050 Povo, Italy
+39-0471-013904; giovanni.modanese@unibz.it

**Abstract.** It has been known for some time that the value $\Lambda$ of the vacuum energy density affects the propagation equation for gravitons - the analogue of photons for the gravitational field. (For historical reasons, $\Lambda$ is also called "cosmological constant".) More precisely, if $\Lambda$ is not zero, then a mass term appears in the propagation equation, such that $m^2=-\Lambda$. As a consequence, the polarization states of gravitons also change, because a massless particle has only two polarization states while a massive particle has more. This effect of the $\Lambda$-term has been confirmed by recent calculations in a curved background, which is actually the only proper setting, since solutions of the classical Einstein equations in the presence of a $\Lambda$-term represent a space with constant curvature. A real value for the mass (when $\Lambda<0$) will show up as a slight exponential damping in the gravitational potential, which is however strongly constrained by astronomical data. The consequences of an imaginary mass (for $\Lambda>0$) are still unclear; on general grounds, one can expect the onset of instabilities in this case. This is also confirmed by numerical simulations of quantum gravity which became recently available. These properties gain a special interest in consideration of the following. (1) The most recent cosmological data indicate that $\Lambda$ is positive and of the order of 0.1 J/m$^3$. Is this value compatible with a stable propagation of gravitons? (2) The answer to the previous question lies perhaps in the scale dependence of the effective value of $\Lambda$. It could then happen that $\Lambda$ is actually negative at the small distance/large energy scale at which the quantum behavior of gravitational fields and waves becomes relevant. Applications for an advanced propulsion scheme is that local contributions to the vacuum energy density (remarkably in superconductors in certain states, and in very strong static electromagnetic fields) can change locally the sign of $\Lambda$, and so affect locally the propagation and the properties of gravitons. The graviton wavefunction, for different values of the parameters, may be characterized by superluminal phase velocity or by unitarity only in imaginary valued time.

## INTRODUCTION

Gravitons are the analogue of photons for the gravitational field, namely the field quanta which are the ultimate constituents of gravitational waves. The recent surge of interest for high frequency gravitational waves (Baker, 2004), in which the single-quantum behaviour play an important role, has prompted new studies of graviton propagation: If high frequency gravitational waves can be successfully generated, possible techniques for deviate or focalize them will be one of the next hot issues.

The propagation of gravitons in flat space-time or in a space-time with small curvature can be described by equations which are similar to those encountered in electromagnetism. In this context, the vacuum energy density plays the role of a very special "medium", which can give gravitons an effective mass. In the following sections we shall consider the cases of zero mass, real mass and imaginary mass. While the zero-mass case is well known and settled at the text-book level, the case of real non-zero mass (refraction index $n>1$, in an optical analogy) has been treated in only few works, with uncertain results (see references in Section B); the case of imaginary mass ($n<1$) is still unexplored.

In this work we shall give a simplified treatment of all the three cases above, based upon the weak-field approximation. The most innovative part of our investigation, however, concerns the possibility of "tuning" locally the vacuum energy density (Section D), and so change the effective refraction index felt by gravitons in their propagation. It turns out that systems with macroscopic quantum coherence, notably superfluids and superconductors, are the best candidates for this purpose.

Note: SI units cannot be used in this matter, because they were too impractical. The employed units ("natural units") are introduced in Section C, where the conversion factors to SI units are also given.

## A. Gravitons in the Linearized Einstein Theory without Vacuum Energy Density

The linearized theory is appropriate to situations where space-time is almost flat and gravitational fields are very weak, i.e. to all situations which do not involve very dense and massive bodies. In this approximation, the field equations in vacuum are linear, and so gravitational field and waves do not self-interact, nor interact with those produced by other sources. The field equations contain only derivatives of the field. Their plane wave solution has the form

$$\Psi = \text{Re}\{h_{\mu\nu} e^{i(\vec{k}\vec{x}-\omega t)}\}. \tag{1}$$

(For an electromagnetic wave one has $\Psi = \text{Re}\{A_\mu e^{i(\vec{k}\vec{x}-\omega t)}\}$, where $A_\mu$ is the four-vector potential.) The constant symmetric tensor $h_{\mu\nu}$ denotes the components of the field, and the exponential its space-time dependence. The wave vector $\vec{k}$ is related to the frequency $\omega$ by $k = |\vec{k}| = 2\pi\lambda^{-1} = \omega/c$. It can be shown (Misner et al., 1970) that the non-vanishing components of $h_{\mu\nu}$ are only two, called the transversal components with helicity $\pm 2$. In the quantum theory we can regard the (complex) $\Psi$ as the wave function of a particle called graviton, having spin 2, momentum $\vec{p} = \eta\vec{k}$ and energy $E = \eta\omega$. It has zero rest mass and its spin can be either parallel to $\vec{p}$ (helicity component +2) or anti-parallel to it (helicity component –2). While photons, which have helicity $\pm 1$, are emitted by oscillating dipoles or by quantum transitions between states whose angular momenta differ by $\Delta l = \pm 1$, gravitons are emitted by oscillating quadrupoles or by quantum transitions between states with $\Delta l = \pm 2$.

In spite of these similarities, gravitons are much more difficult to observe than photons. The probability that a high-frequency graviton is emitted in an atomic or nuclear transition is generally small, because the coupling constant $G/c^4$ is much smaller than the electromagnetic coupling $e^2/\varepsilon_0\eta c$. So it is more likely that a transition between states with $\Delta l = \pm 2$ generates two photons (unless the photon production is forbidden for some reason, or the graviton production amplified (Fontana and Baker, 2003)).

The main natural sources of gravitons are then astrophysical systems like binary stars, collapsing or exploding stars with large mass and variable quadrupolar momentum. The corresponding frequencies are low, up to 1 KHz max, so the single gravitons in the emitted gravitational waves carry a very small energy and cannot be easily observed.

The $1/r^2$ behavior of the static gravitational force is also a consequence of the fact that the graviton has zero mass. This can be deduced from the classical field equations, or in the quantum theory, where the force can be thought as generated by the exchange of virtual gravitons, in (partial) analogy to the electromagnetic and the weak and strong forces of the standard model (Modanese, 1995).

## B. Gravitons with Real Mass

If the wave equation of a field/particle contains, besides the derivatives of the field, also a positive term $m^2\Psi$ proportional to the field itself (see the next section for the explicit form of this equation), then the solution is still of the plane-wave form $\Psi \sim e^{i(\vec{k}\vec{x}-\omega t)}$, but

(1) The relation between $k$ and $\omega$ is different; the phase velocity $v$ depends on $\omega$ and is always less than $c$. We have a "dispersion relation"

$$v(\omega) = \frac{k}{\omega} = c\sqrt{1 - \frac{m^2 c^4}{\eta^2 \omega^2}} \ . \tag{2}$$

(2) All the spin components are different from zero. For instance, the wave function of a massive spin 2 particle has 5 components, with helicities $0, \pm 1, \pm 2$.

(3) The static force carried by virtual particles behaves like $e^{-mcr/\eta}/r^2$ (Yukawa potential). From astrophysical observations of the Newton force, the Particle Data Group is currently quoting an experimental limit of $m_{graviton} < 10^{-38} m_{nucleon}$, corresponding to a very large Yukawa range, namely $r \sim 6 \cdot 10^{22}$ m.

Now, starting from the Einstein equations in vacuum there is only one way, consistent with general covariance (i.e., independent on the coordinate system) to introduce a mass term: one adds a source term of the form $T_{\mu\nu} = \frac{c^4 \Lambda}{8\pi G} g_{\mu\nu}$, where $\Lambda$ is a constant; the quantity $\frac{c^4 \Lambda}{8\pi G}$ is called vacuum energy density, because it represents a sort of homogeneous, isotropic and Lorentz-invariant background (i.e. it is the same for all observers, independent from their relative motion). We shall see later what the origin of this background could be. In the weak field approximation the field equations then give a solution with mass such that $m^2 = const. \cdot \Lambda$, where "$const.$" is a positive, purely numerical factor. For $\Lambda > 0$, the graviton acquires a real mass $m \sim \frac{\eta}{c}\sqrt{\Lambda}$ (Veltman, 1976). Note that the signs in the equations above are valid for metric signature $\eta_{\mu\nu} = (1,-1,-1,-1)$, otherwise the sign of $\Lambda$ is reversed.

Beyond the linearized approximation, things get much more complicated. The vacuum energy background gives space-time a small constant curvature, so one has to re-analyse all particle properties in the context of curved space. It is difficult to distinguish, in the solutions of field equations, the effects of a mass term from those of the space-time curvature, because both manifest themselves over (large!) distances $r \sim 1/\sqrt{\Lambda}$. It is not surprising that there is no general agreement in the literature about the properties of gravitons in these circumstances. The relation between $m$ and $\Lambda$ is confirmed and specified to be $m^2 = \frac{2}{3}\Lambda$, but the dispersion relation $\omega/k$ and the number of spin components turn out to be different, depending on the computational approach and the coordinate system employed (Novello and Neves, 2003; Tsamis and Woodard, 1992).

## C. Gravitons with Imaginary Mass

What happens if the mass term $m^2\Psi$ in the field equation is negative, i.e. $m$ is imaginary? The usual answer is that in this case the quantum theory "does not exist", because it is neither stable nor unitary. In order to understand better this point, one needs to write the field equation explicitly. It is quite convenient for this to introduce the so-called "natural" units, chosen in such a way that $\eta = c = 1$. In these units, velocity is adimensional and expressed as a fraction of the velocity of light. As a consequence, space and time have the same unit, the cm; but while as length unit the cm corresponds to the usual cm, as time unit it corresponds to the time it takes for light to travel for 1 cm, i.e. ~ 1 sec/3·10$^{10}$. The unit for energy is the cm$^{-1}$, with the conversion 1 cm$^{-1}$~3·10$^{-23}$ J. Mass is also measured in cm$^{-1}$, with the conversion 1 cm$^{-1}$~10$^{-37}$ g. The Newton constant $G$ has dimension $l^2$ and is of the order of 10$^{-66}$ cm$^2$.

The wave equation in these units is

$$\left(\frac{\partial^2}{\partial t^2} - \nabla^2\right)\Psi + m^2\Psi = 0.$$  (3)

Substituting a function with the form of a plane wave

$$\Psi = \Psi_0 e^{i(\vec{k}\vec{x} - \omega t)} = \Psi_0 e^{i(\vec{p}\vec{x} - Et)},$$  (4)

then it is a solution if

$$E^2 - p^2 = m^2.$$  (5)

If $m^2 > 0$ (real mass), expressing $E$ in terms of $p$ one finds for the positive energy solutions $E = \sqrt{p^2 + m^2}$.

If $m^2 < 0$ (imaginary mass), we can write $E^2 = p^2 - |m^2|$ and there are two possible cases:

(1) For large impulses, such that $p^2 > |m^2|$, we have $E^2 > 0$ and $E$ is real. In this case, the wave function is still a plane wave periodic in time, with superluminal phase velocity

$$v = \frac{p}{|E|} = \sqrt{1 + \frac{|m^2|}{E^2}}.$$  (6)

The graviton would thus become a 'tachyon", a bizarre kind of particle that has long been postulated but never observed (Rembielinski, 1992, and ref.s). According to special relativity, the propagation of tachyons violates causality relations, and so they are usually thought to play only a marginal role in some cosmological models and in string theory.

(2) For small impulses, such that $p^2 < |m^2|$, we have $E^2 < 0$ and $E$ is imaginary. In this case, the wave function still has an oscillating behavior in space, but changes in time as a real exponential:

$$\Psi = \Psi_0 e^{i\vec{p}\vec{x}} e^{\pm Et}.$$  (7)

Therefore the squared module of $\Psi$, giving the probability to find the particle, is not constant in time – which is also called the "non-unitarity" property. Alternatively, the wave function is found to be unitary in imaginary valued time. It is hard to believe that a particle could fade out of reality as a function of time, instead the possible existence of imaginary time is consistent with the approach followed in Euclidean quantum gravity (Gibbons, 1993) and theories on Faster-Than-Light travel in higher dimensional spacetimes (Froning, 2004).

## D. Global and Local Values of the Vacuum Energy Density

In this Section we shall give the sign and magnitude order of $\Lambda$ according to the cosmological observations, i.e. at very large scale, and the postulated sign and magnitude order of possible local contributions to the vacuum energy density due to the presence of certain physical systems (static electromagnetic fields, superconductors). Since

$m^2_{graviton} = \frac{2}{3}\Lambda$, the sign and magnitude of $\Lambda$ define whether the graviton mass is real or imaginary, and how much infra- or super-luminal is its propagation, and for which values of its momentum, and how quickly, its wave function changes in the case of imaginary energy.

The latest measurements of the expansion rate of the universe yield $\Lambda \sim 10^{-50}$ cm$^{-2}$ as most probable value of the cosmological constant, which implies a vacuum energy density, in SI units, $\Lambda/8\pi G \sim 10^{-1}$ J/m$^3$. Earlier observations set an upper limit on $\Lambda$ of the order of $10^{-54}$ cm$^{-2}$, so it was thought to vanish exactly for symmetry reasons. In fact, the observed non-zero value of $\Lambda$ creates an arduous fine-tuning problem, because quantum fields give large contributions with different signs to the cosmological constant, in two ways:

(1) Their zero-point oscillations are associated with a huge energy density, whose value depends on the frequency cut-off.

(2) Some fields can have a non-vanishing vacuum expectation value $\Phi_0 = <0|\Phi|0>$. Such contributions are essentially classical. If *L* is the field lagrangian, the corresponding cosmological term for a scalar field is $-8\pi GL(\Phi_0)$. In elementary particle physics a non-vanishing vacuum expectation value is usually the consequence of a spontaneous symmetry breaking process.

The above mentioned value of $\Lambda$ is global, in the sense that it is supposed to define a uniform background present in the whole universe. A local contribution to the vacuum energy density can arise when the state of a localized physical system is described by a classical field comparable with the vacuum expectation value of a quantum field. Some interesting cases of this kind occur in condensed matter physics. In this context, the physical systems properly described by continuous classical-like fields (also at microscopic level, not just in a macroscopic-average sense as for fluids) are basically the following:

- The electromagnetic field in the low-frequency limit, in states where the photons number uncertainty is much larger than the phase uncertainty. In this case, the vacuum energy contribution is equal to the part of the energy-momentum tensor proportional to $g_{\mu\nu}$, i.e. to ($B^2$-$E^2$). The magnitude order of this contribution is, for instance, $\sim 10^2$ J/m$^3$ for an electrostatic field $E \sim 10^6$ V/m and $\sim 10^5$ J/m$^3$ for a static magnetic field $B \sim 1$T.

- Systems with macroscopic quantum coherence, described by "order parameters", like superfluids, superconductors and spin systems. For a superconductor described by a collective wave function obeying the Ginzburg-Landau equation, in the non-relativistic limit it is possible to express *L* as a function of $|\psi_{GL}|$ (Modanese, 2003). Relevant values of *L* are those for the case of constant pairs density (for instance, *L* up to $\sim 10^6$ J/m$^3$ for YBCO) and those at local density extrema (|*L*| up to $\sim 10^8$ J/m$^3$ for YBCO; *L*>0 for maxima, *L*<0 for minima).

Let us now see which graviton mass corresponds to the cosmological value of $\Lambda$ and to the different kinds of local "effective $\Lambda$". To the cosmological value $\Lambda \sim -10^{-50}$ cm$^{-2}$ corresponds an imaginary mass with absolute value $|m| \sim 10^{-25}$ cm$^{-1} \sim 10^{-62}$ g. According to our previous discussion, and taking advantage of the simple properties of the natural units, we find that gravitons with wavelength $\lambda < 10^{25}$ cm have real energy and velocity exceeding the velocity of light by a factor

$$v = \sqrt{1 + \frac{|m|^2}{E^2}} = \sqrt{1 + |m|^2 \lambda^2} \sim \sqrt{1 + 10^{-50}\lambda^2} \ . \quad (8)$$

Gravitons with wavelength $\lambda > 10^{25}$ cm (i.e., larger than $10^7$ light years!) have imaginary energy and the absolute value of their wave function evolves exponentially with a time constant of $\sim 10^{25}$ cm $\sim 10^{15}$ sec $\sim 10^9$ years.

The negative effective $\Lambda$ in superconductors can be $\sim 10^9$ times larger than the cosmological value (compare the estimate of the vacuum energy density $\Lambda/8\pi G$ above), but it still very small. We then have $|m| \sim 10^{-20}$ cm$^{-1} \sim 10^{-57}$ g. Gravitons with wavelength $\lambda < 10^{20}$ cm have real energy and $v \sim \sqrt{1 + 10^{-40}\lambda^2}$. (The case $\lambda > 10^{20}$ cm is not applicable here.) In the case of positive effective $\Lambda$ with the same magnitude order, gravitons will have a small real mass $m\sim10^{-20}$ cm$^{-1}$ and propagation velocity slightly infra-luminal $v \sim \sqrt{1-10^{-40}\lambda^2}$. The static gravitational force will have a Yukawa range of the order of $10^{20}$ cm.

The possibility of Faster-Than-Light travel is based on a class of postulates regarding the significance of the speed of light, the validity of Special Relativity and General Relativity near and beyond *c*, the possibility of travelling in imaginary valued time, etc. Neither science fiction nor current scientific literature clearly explains "what" can really do that "jump in hyperspace" that will enable the exploration of the galaxy. Nevertheless if an object or a particle is found to be characterized by some suitable properties, then Faster-Than-Light travel is no longer science fiction, but a fact. Studying graviton propagation in spacetimes with a cosmological constant we have found superluminality or unitarity of the graviton wavefunction only in imaginary time. Imaginary valued time is equivalent to a spatial coordinate, therefore under these circumstances the graviton is somehow locked in time and propagates in a space with 4 space-like coordinates. These findings open new questions; for instance, if imaginary time is required to keep the probability to find the particle constant in time, does imaginary time really exist in the physical world? We expect that experiments should satisfy our curiosity in the near future, as soon as coordinated generators and detectors of gravitational waves will be made available.

## CONCLUSIONS

Our results add to the increasing body of knowledge about possible generation and exploitation of high frequency gravitational waves for data transmission and space propulsion.

Standard field-theoretic arguments like those of the previous Sections strictly apply to the case when the graviton mass is constant over all space. We have seen, however, that the vacuum energy density can change locally. In a realistic model one will have to take this possibility into account, and consider for instance the propagation of gravitons which are generated in a region where they have real mass and then enter a region where their mass is imaginary; and even within such regions, the value of the mass can change during propagation. A formal treatment of such situations in quantum field theory is clearly very difficult. From the physical point of view, however, they are nothing extravagant: just think of the analogy to light propagation through media with variable optical properties. In order to obtain a viable model, one will therefore have to resort to an empirical/phenomenological approach. It is perhaps possible to consider first a semi-classical approximation, with the graviton propagating in the "variable -mass potential" $m^2(x)\Psi^2(x)$.

If we want to describe graviton propagation in vacuum over distances of the order of the laboratory scale (m or km) we need to know the effective value of $\Lambda$ at this scale. This is possibly different from the cosmological value, which is an average over very large distances (Shapiro and Solà, 2000). Remember that the value of $\Lambda$ at very short distances depends on the high energy behavior of strong and electro-weak interactions, on the onset of super-symmetry etc.; in much the same way, the value of $\Lambda$ at scales about 1 m – 1 km depends on the existence of very light particles, with mass of the order of $10^{-40}$ g, which are still unknown (they are possible constituents of the dark matter, interacting very weakly with ordinary matter).

The perturbative treatment of graviton propagation outlined in this paper has some inherent limitations. In particular, in the case of imaginary mass a full strong-field treatment could well give evidence of more dramatic instabilities ("dipolar fluctuations" (Modanese, 2003)) a nd non-unitary behavior. For instance, in the numerical simulations of Euclidean quantum gravity by Hamber (2000), discretized spacetimes with $\Lambda<0$ quickly "collapse", because the configurations with small volume are energetically more favourable in that case. The magnitude orders given in this paper are therefore quite conservative.

In conclusion, we have found that certain physical systems behave, with respect to graviton propagation, like the analogue of optical media with refraction index $n<1$ and $n>1$. This happens when their local contribution to the vacuum energy density is respectively positive or negative, and so the induced graviton mass is respectively real or imaginary.

We expect that "*n*" can differ substantially from 1 in the case $n>1$, due to non-perturbative quantum gravity effects. (There is also some experimental evidence for this, because the gravitational anomalies in superconductors

(Podkletnov, 1997; Podkletnov and Modanese, 2003) seem to require a negative vacuum energy density (Modanese, 2003)).

It is therefore conceivable that "gravitational lenses" can be built in the laboratory, resembling those observed in astrophysics, but such to offer a choice of different shapes and effects. This represents the main practical conclusion of this work, relevant for possible applications to space propagation or data transmission, and a strong motivation for continuing our investigation. We can imagine that artificially-generated gravitons can be driven through such lensing devices in order to obtain directional thrust. Or we can imagine to use lensing devices to collect and focalize natural gravitons from an homogeneous cosmic background which would otherwise be inaccessible. In addition, because gravitons have positive mass near their source, a conceptually simple propulsion system based on reaction can be conceived, with the great advantage that the mass of the graviton is not a part of spacecraft mass, the propulsion system is based on reaction but it is not a rocket.

Our formulation also suggests the possible physical existence of imaginary valued time. This discovery is surprising and requires experimental investigation, which could be performed with coordinated generators and detectors of gravitational waves.

## NOMENCLATURE

$E$, $p$, $v$, $\omega$, $\lambda$ = graviton energy, momentum, velocity, frequency, wavelength

$k$ = graviton wave-number, $k=2\pi\lambda^{-1}$

$m$ = graviton mass

$\Psi$ = graviton wave-function

$\Lambda$ = "cosmological constant"; the vacuum energy density is equal to $c^4\Lambda/8\pi G$

$T_{\mu\nu}$ = energy-momentum tensor of matter

$g_{\mu\nu}$ = metric tensor

$\eta_{\mu\nu}$ = metric tensor of flat space-time: $\eta_{\mu\nu}$ = diag(1,-1,-1,-1)

$L$ = lagrangian density of matter

$\psi_{GL}$ = wave-function of Cooper pairs in the Ginzburg-Landau model

## ACKNOWLEDGMENTS

G.F. wishes to acknowledge Dr. Franklyn Mead and AFRL for a grant to attend the STAIF 2004 Conference.